\documentclass[aps,prl,twocolumn,superscriptaddress,letterpaper,longbibliography]{revtex4-2}
\usepackage{amssymb,amsmath,amsthm}
\usepackage[colorlinks=true, linkcolor=blue, citecolor=blue, urlcolor=blue]{hyperref}
\usepackage{graphicx}
\usepackage{physics}
\usepackage{soul}

\begin{document}

\title{Topological $\pi/2$ modes in photonic waveguide arrays}

\author{Gang Jiang}
\affiliation{National Laboratory of Solid State Microstructures and Department of Physics, Nanjing University, Nanjing 210093, China}

\author{Siyuan Zhang}
\affiliation{Department of Physics, The Chinese University of Hong Kong, Shatin, Hong Kong SAR, China}

\author{Weiwei Zhu}
\email{phyzhuw@ouc.edu.cn}
\affiliation{College of Physics and Optoelectronic Engineering, Ocean University of China, Qingdao 266100, China}
\affiliation{Key Laboratory for Optics Photoelectronics, Qingdao 266100, China}
\affiliation{Engineering Research Center of Advanced Marine Physical Instruments and Equipment of MOE, Qingdao 266100, China}

\author{Y. X. Zhao}
\affiliation{Department of Physics and HKU-UCAS Joint Institute for Theoretical and Computational Physics at Hong Kong, The University of Hong Kong, Pokfulam Road, Hong Kong, China}
\affiliation{HK Institute of Quantum Science \& Technology, The University of Hong Kong, Pokfulam Road, Hong Kong, China}

\author{Haoran Xue}
\email{haoranxue@cuhk.edu.hk}
\affiliation{Department of Physics, The Chinese University of Hong Kong, Shatin, Hong Kong SAR, China}
\affiliation{State Key Laboratory of Quantum Information Technologies and Materials, The Chinese University of Hong Kong, Shatin, Hong Kong SAR, China}

\begin{abstract}
Periodic driving is a powerful tool to generate exotic topological phases without static counterparts, such as the anomalous chiral edge modes from bulk bands with zero Chern number and topological $\pi$ modes exhibiting period-doubled dynamics. Recently, a new class of Floquet topological mode, namely the $\pi/2$ mode, which carries four-period periodicity and has potential applications in quantum computing, was proposed based on a square-root method and realized in an acoustic system. Here we propose a laser-written waveguide array lattice to realize topological $\pi/2$ modes in photonics. Our photonic model simulates a square-root periodically driven Su-Schrieffer-Heeger model and has a rich phase diagram allowing for the co-existence of conventional zero, $\pi$ modes, and the new $\pi/2$ modes. Through numerical simulations of the wave equation, we uncover the unique four-period evolution feature of the $\pi/2$ modes. Our model, which only contains four waveguides per unit cell and two driving steps, is easy to implement with current fabrication techniques and may find applications in quantum optics.
\end{abstract}

\maketitle

\textit{Introduction.}---
Periodic driving, also known as Floquet engineering, has gained significant research interest in the study of topological phases over the past decade~\cite{oka2019floquet, rudner2020band}. Floquet bands are periodic not only along the momentum axis but also along the energy axis, making the topological properties of Floquet systems essentially different from the static ones. In particular, Floquet engineering can generate unique topological modes that are absent in static systems.  A well-known example is the anomalous Floquet topological insulator with zero Chern numbers for the bulk bands but still hosts chiral edge modes at its boundary~\cite{rudner2013anomalous}. These chiral edge modes are characterized by a Floquet winding number and exhibit superior robustness over conventional chiral edge modes in Chern insulators~\cite{zhang2021superior}. Another example is the topological mode with a quasienergy of  $\pi/T$ with $T$ the driving period (referred to as $\pi$ mode hereafter), which exhibits rich dynamic features such as time-dependent intensity distribution and period-doubled evolution~\cite{jiang2011majorana, cheng2019observation, zhu2022time}. Owing to their unique properties, these Floquet topological modes are candidates for various applications, including quantum computing and robust waveguiding~\cite{oka2019floquet, rudner2020band, zhang2021superior}.

With the rapid progress in topological photonics and acoustics~\cite{ozawa2019topological, xue2022topological}, various Floquet topological modes have been implemented for light and sound, including the above-mentioned anomalous chiral edge modes and $\pi$ modes~\cite{rechtsman2013photonic, gao2016probing, peng2016experimental, maczewsky2017observation, mukherjee2017experimental, cheng2019observation, maczewsky2020fermionic, zhang2021superior, wu2021floquet, zhu2022time, sidorenko2022real, cheng2022observation}. Moreover, the high flexibility of synthetic platforms makes it possible to realize more complicated Floquet models. Recently, a new type of Floquet topological mode with a quasienergy of $\pi/2T$, namely the $\pi/2$ mode, was discovered, which can lead to a surprising 4$T$-periodic evolution~\cite{sreejith2016parafermion, bomantara2021z}. To realize this mode in wave systems, a square-root procedure is proposed, which transforms a two-band Floquet Su-Schrieffer-Heeger (SSH) model into a four-band one~\cite{bomantara2022square}. So far, the $\pi/2$ mode has only been observed in an acoustic lattice~\cite{cheng2022observation}.

\begin{figure*}
    \includegraphics[width=1.0\textwidth]{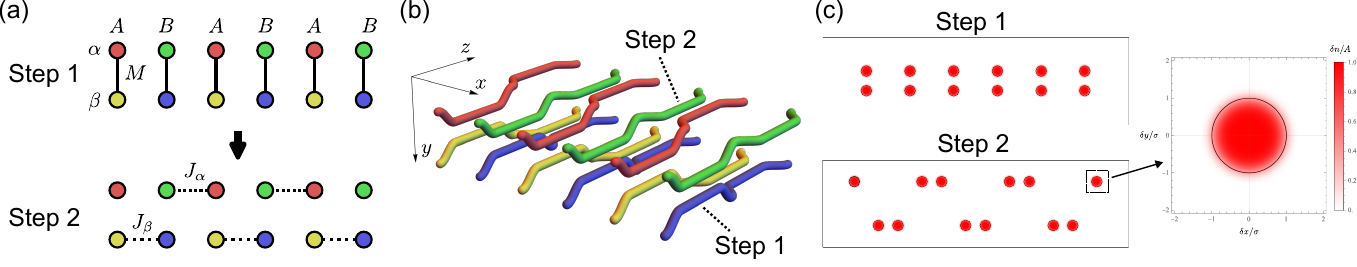}
    \caption{Square-root Floquet SSH model and its photonic realization. (a) The two driving steps of the square-root Floquet SSH model. (b) Schematic sketch of the photonic system realizing the lattice model in (a). The coupling between two adjacent waveguides is enabled when they are close to each other. (c) Refractive index distribution in the two driving steps. The right inset shows refractive index modulation $\delta n$ of a waveguide and $(\delta x,\delta y)$ is the position vector in the $xy$-plane with respect to the center of the waveguide. The radius of the black circle is $\sigma$. } 
    \label{fig:lattice}
\end{figure*}

In this work, we propose a concrete design to realize the $\pi/2$ mode in a photonic system. Specifically, we design a laser-written waveguide lattice that can be mapped to the square-root Floquet SSH tight-binding model. The required driving is realized by bending the waveguides along the evolution direction. By adjusting the distance between the waveguides, we can tune the corresponding coupling coefficients and realize four topologically distinct phases, including a trivial phase, a topological phase with zero and $\pi$ modes, a topological phase with $\pm\pi/2$ modes, and a topological phase with zero, $\pi$ and $\pm\pi/2$ modes. Using simulations of the wave equation, we then study the field evolution under a single-site excitation, closely mimicking the experimental measurements. A 4$T$-periodic evolution is observed when zero, $\pi$ and $\pm\pi/2$ modes are simultaneously present. Compared with the evolution dynamics of the other three phases, the existence of the $\pi/2$ modes can be clearly identified. We note that laser-written waveguide arrays have been substantially developed in recent years to study various topological phases~\cite{kang2023topological}. Thus, our model is ready to be tested in experiments. 

\textit{Square-root Floquet SSH model.}---
A Floquet system is defined by a unitary evolution operator $U$ over one driving period. In general, the one-period evolution operator, also known as the Floquet operator, can be written as 
\begin{equation}
    U = \mathcal{T} \exp(-\mathrm{i} \int_{0}^{T} H(t) \,dt )
    , \label{eq:floquet}
\end{equation} 
where $H(t)$ is the Hamiltonian of the Floquet system with the period $T$ (i.e., $H(t+T)=H(t)$) and $\mathcal{T}$ is the time-ordering operator. The eigenvalues of $U$ are given as $e^{-\mathrm{i} \epsilon T}$, where $\epsilon$ is the so-called quasienergy with a period of $2\pi/T$. 

\begin{figure*}[ht]
    \includegraphics[width=0.90\textwidth]{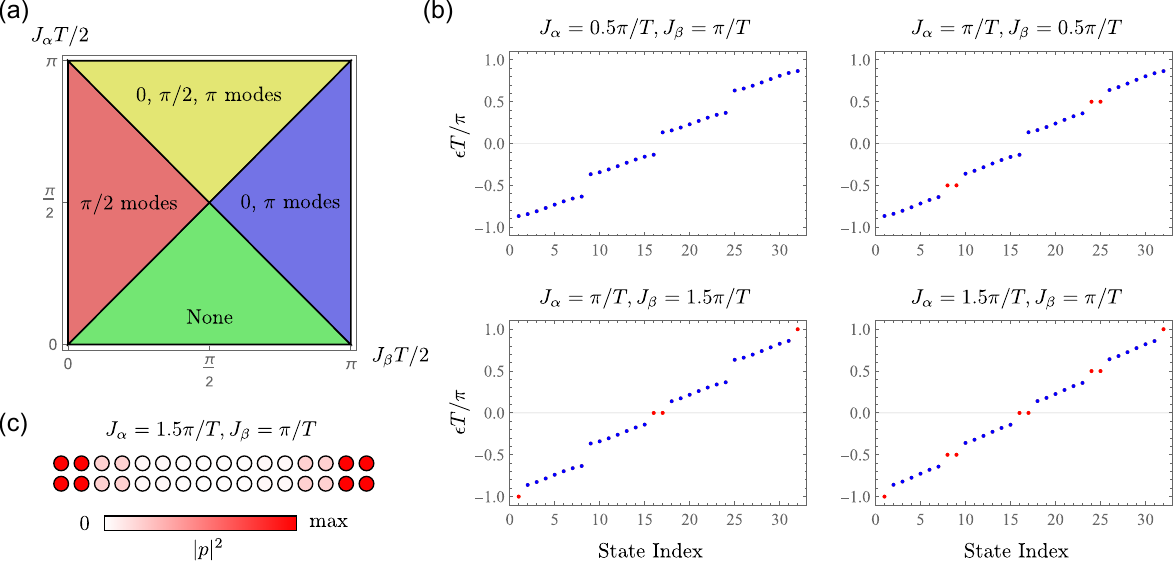}
    \caption{Eigenspectra of the lattice model. (a) Phase diagram for the square-root Floquet SSH model at $M=\pi/T$. The four distinct phases are highlighted in different colors. (b) Representative quasienergy spectra for the four phases, with topological edge modes plotted in red. We use a finite system with 32 sites and take $M=\pi/T$ in the calculation. (c) Eigenmode profiles for the $\pm \pi/2$ modes in the yellow region (i.e., $M=\pi/T$, $J_{\alpha}=1.5\pi/T$ and $J_{\beta}=\pi/T$). 
    \label{fig:spectrum}}
\end{figure*}

Consider a square-root Floquet SSH model described by the following two-step driving protocol:
\begin{equation}
    {H}^{\text{SR}}\left( t \right) = \left\{ {
        \begin{aligned}
        h^{\text{SR}}_1 & \qquad nT < t \leqslant \left(n + \frac{1}{2}\right)T \\ 
        h^{\text{SR}}_2 & \qquad \left(n + \frac{1}{2}\right)T < t \leqslant \left( {n + 1} \right)T 
        \end{aligned}
        } 
    \right. , \label{eq:hami_SR_0}
\end{equation} 
where
\begin{align}
    h^{\text{SR}}_1 =& \sum_{j=1}^N \sum_{S=A,B} M|j,S,\alpha\rangle \langle j,S,\beta | +h.c.  , 
    \label{eq:hami_SR_1} \\
    h^{\text{SR}}_2 =& \sum_{j=1}^{N-1} J_\alpha  |j,B,\alpha\rangle \langle j+1,A,\alpha | 
    \nonumber  \\
    &+ \sum_{j=1}^{N} J_\beta |j,A,\beta \rangle \langle j,B,\beta | +h.c..
    \label{eq:hami_SR_2}
\end{align}
Here, $M$, $J_\alpha$ and $J_\beta$ are coupling coefficients, $|j,S,\xi\rangle$ denotes a state at sublattice $S=A,B$ of the $j$th site in the chain species $\xi=\alpha,\beta$, and $n \in \mathbb{Z}$, as illustrated in Fig.~\ref{fig:lattice}(a). This model was previously realized using an acoustic crystal with tailored connections between acoustic waveguides~\cite{cheng2022observation}. In this work, we propose a photonic realization based on laser-written optical waveguide arrays, where the propagation direction (i.e., the $z$ axis) plays the role of time. The time-dependent couplings are achieved by controlling the spacing between adjacent waveguides [see Fig.~\ref{fig:lattice}(b) and (c)]. Note that, unlike the acoustic realization where all waveguides are straight~\cite{cheng2022observation}, here we need to bend the waveguides to realize desired couplings, which is feasible using the current laser-written technique~\cite{rechtsman2013photonic, maczewsky2017observation, mukherjee2017experimental, maczewsky2020fermionic}.

\begin{figure}[b]
    \includegraphics[width=0.90\columnwidth]{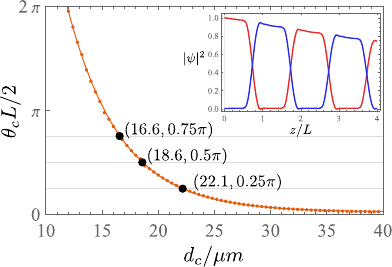}
    \caption{Correspondence between the coupling $\theta_c$ and the waveguide separation $d_c$. The dots are numerically obtained data and the curve is an exponential fit. The inset shows the intensity evolution in the two waveguides whose separation is $18.6$ $\mathrm{\mu m}$.
    \label{fig:fit}}
\end{figure}

\begin{figure*}
    \includegraphics[width=0.90\textwidth]{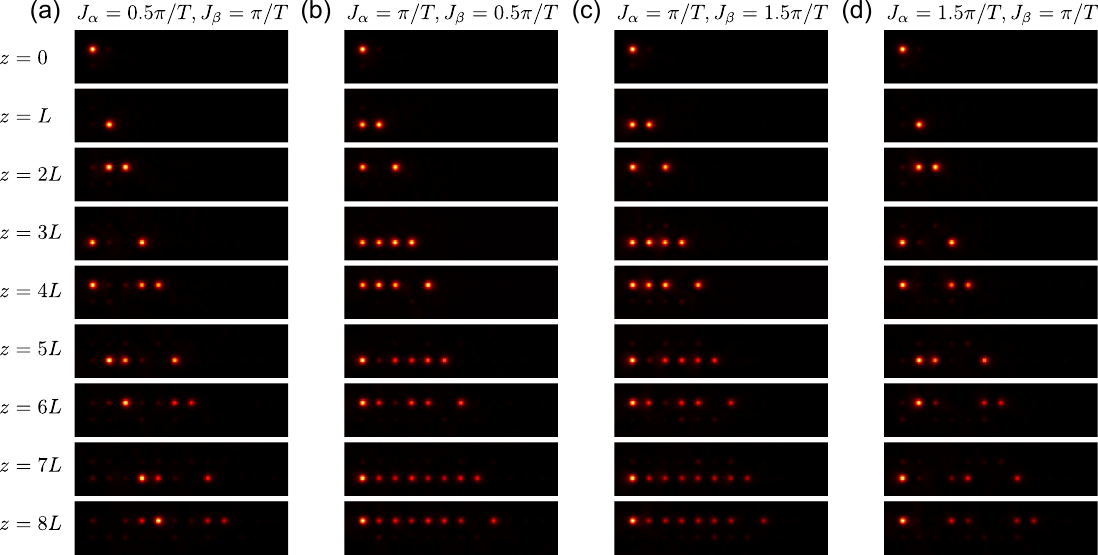}
    \caption{Mode evolution of the simulation with different parameters $J_\alpha$ and $J_\beta$, and $M=\pi/T$. The initial excitation is put at the site in the top left corner. (a) Trivial phase without periodicity. (b) Topological phase with $\pm \pi/2$ modes with $2L$ periodicity. (c) Topological phase with $0$ and $\pi$ modes with $2L$ periodicity. (d) Topological phase with $0$, $\pi$, and $\pm \pi/2$ modes with $4L$ periodicity.
    \label{fig:evolution}}
\end{figure*}

This lattice model can be understood as a nontrivial square root of a Floquet SSH model~\cite{bomantara2022square, arkinstall2017topological}. The Floquet SSH model (hereafter referred to as the parent model) describes a system of a single 1D chain subjected to a two-step driving: 
\begin{equation}
    {H}^{\text{P}}\left( t \right) = \left\{ {
        \begin{aligned}
        h^{\text{P}}_1 & \qquad nT < t \leqslant \left(n + \frac{1}{2}\right)T \\ 
        h^{\text{P}}_2 & \qquad \left(n + \frac{1}{2}\right)T < t \leqslant \left( {n + 1} \right)T 
        \end{aligned}
        } 
    \right. , \label{eq:hami_P_0}
\end{equation} 
where
\begin{align}
    h^{P}_1 =& {J_\alpha }|j,B\rangle \langle j+1,A | +h.c.\;,
    \label{eq:hami_P_1} \\
    h^{P}_2 =& {J_\beta }|j,A\rangle \langle j,B | +h.c.\;.
    \label{eq:hami_P_2}
\end{align}
The Floquet operator of the square-root model and the parent model are 
\begin{align}
    U^{\text{SR}} =& U_2^{\text{SR}} U_1^{\text{SR}}= \exp\left(-\mathrm{i} h_2^{\text{SR}}\frac{T}{2}\right) \exp\left(-\mathrm{i} h_1^{\text{SR}}\frac{T}{2}\right) ,
    \label{eq:evolution_SR} \\
    U^{\text{P}} =& U_2^{\text{P}} U_1^{\text{P}}= \exp\left(-\mathrm{i} h_2^{\text{P}}\frac{T}{2}\right) \exp\left(-\mathrm{i} h_1^{\text{P}}\frac{T}{2}\right),
    \label{eq:evolution_P}
\end{align}
respectively. 
Mathematically, at $MT = (2m+1)\pi$ with $m\in \mathbb{Z}$, the two Floquet operators have the relationship 
\begin{align}
    {U^\text{SR}} = \left( {
        \begin{array}{*{20}{c}}
            0&{ - \mathrm{i} U_1^\text{P}} \\ 
            { - \mathrm{i} U_2^\text{P}}&0 
        \end{array}} 
    \right) 
    , \label{eq:evolution_relationship_0}
\end{align}
and 
\begin{equation}
    \left( U^\text{SR} \right)^2  = -\text{diag} \left( U_1^\text{P} U_2^\text{P}, U_2^\text{P} U_1^\text{P} \right)
    . \label{eq:evolution_relationship_1}
\end{equation} 
From Eq.~\eqref{eq:evolution_relationship_1}, we obtain that as $MT$ equals to $(2m+1)\pi$, $\left( U^\text{SR} \right)^2$ is directly related to $U^\text{P}$. Even if the $MT$ deviates little from $(2m+1)\pi$, $\left( U^\text{SR} \right)^2$ still exhibits the physics expected from its parent system because of the robustness of Floquet phases~\cite{bomantara2022square}. 

Depending on the system parameter values, the square-root model hosts different topological edge modes, as shown in Fig.~\ref{fig:spectrum}(a) and (b). The green and blue regions support no topological edge modes, and zero and $\pi$ modes, respectively. These two regions also exist in the original phase diagram of the parent model. In the other two regions (i.e., the red and yellow regions), by contrast, there emerge the desired $\pi/2$ modes due to the square-root procedure. 

The topological $\pi/2$ modes can be characterized by the topological invariant of the parent model associated with the zero gap~\cite{bomantara2022square, cheng2022observation}:
\begin{equation}
    \nu^{\text{P}}_0 = \frac{1}{ 2 \pi \mathrm{i} } \int_{-\pi}^{\pi} \frac{1}{z_0 + z'_0} \frac{d z_0 }{d k} d k,
    \label{eq:ti_p_0} \\
\end{equation}
where 
\begin{align}
    z_0 =& \sin(\frac{J_\alpha T }{4})\cos(\frac{J_\beta T }{4}) e^{ \mathrm{i} k} ,\\
    z'_0 =& \cos(\frac{J_\alpha T }{4})\sin(\frac{J_\beta T }{4}).
\end{align}
It can be found that when $ \left| \sin(\frac{J_\alpha T }{4})\cos(\frac{J_\beta T }{4}) \right| > \left| \cos(\frac{J_\alpha T }{4})\sin(\frac{J_\beta T }{4}) \right| $, $\nu^{\text{P}}_0=1$,  which leads to the zero modes in the parent model and subsequently the $\pi/2$ modes in the square-root model. 

\textit{Realization in photonic waveguides.}---Now, we discuss the realization of the square-root Floquet SSH model in the photonic lattice shown in Fig.~\ref{fig:lattice}(b). 
Such a photonic lattice is described by a Schr\"odinger-like equation:
\begin{equation}
    \mathrm{i} \partial_z \psi = -\frac{1}{2k_0} \nabla_{\perp}^2 \psi - \frac{k_0 \delta n (x,y,z)}{n_0} \psi, \label{eq:schrodinger}
\end{equation}
where $\nabla_{\perp}^2 = \partial_x^2 + \partial_y^2$, $k_0 = 2 \pi n_0 / \lambda$ is the wavenumber, $\psi$ is the field distribution, $n_0$ is the background refractive index and $\delta n$ is the refractive index detuning. For concreteness, we set the refractive index as $n_0 = 1.473$ at wavelength $\lambda = 1550$ nm. The refractive index modulation near the waveguide is described by 
\begin{equation}
    \delta n=A e^{-{[(\delta x/\sigma)^2+(\delta y/\sigma)^2]}^3}, \label{eq:modulation}
\end{equation}
where $A=2.6\times 10^{-3}$ and $(\delta x,\delta y)$ is the position vector in the $xy$-plane with respect to the center of the waveguide. This modulation gives rise to a waveguide with a circular cross-section of radius $\sigma=4.9$ $\mathrm{\mu m}$ [see Fig.~\ref{fig:lattice}(c) for visualization for the refractive index distribution]. Each waveguide supports a tightly confined $s$ mode that is isotropic in the $xy$ plane [see Fig.~\ref{fig:evolution}]. The wave equation is numerically solved using the split-step Fourier method~\cite{okamoto2006fundamentals}. In the simulation,  eight driving periods are used, with one-period length $L=20$ mm.

To ensure that we can put the photonic lattice in a desired point in the phase diagram, we first perform simulations of two coupled waveguides to get the suitable separations for different coupling coefficients. 
The ideal coupling we want to realize is described by 
\begin{equation}
    \theta = \left\{ {
        \begin{aligned}
            0 & \qquad nL < z \leqslant (n+0.5)L \\ 
            \theta_c & \qquad (n+0.5)L < z \leqslant (n+1)L \\
        \end{aligned}
        } 
    \right. . \label{eq:coupling}
\end{equation}
Based on the ideal coupling, we make the two waveguides bend towards each other. The designed waveguide separation in one Floquet period is given by 
\begin{equation}
    d = \left\{ {
        \begin{aligned}
            a & \qquad 0 < z \leqslant 0.5L \\ 
            a-\frac{1+\cos\phi}{2}(a-d_c) & \qquad 0.5L < z \leqslant 0.6L \\
            d_c & \qquad 0.6L < z \leqslant 0.9L \\
            a-\frac{1-\cos\phi}{2}(a-d_c) & \qquad 0.9L < z \leqslant L
        \end{aligned}
        } 
    \right. , \label{eq:separation_1}
\end{equation}
where $\phi=10\pi z/L $, $a$ is the initial waveguide separation ensuring negligible coupling ($a=40$ $\mathrm{\mu m}$), and $d_c$ is a tunable separation parameter to realize to a targeted $\theta_c$.

We perform simulations to obtain the field evolutions in the two waveguides, from which the coupling can be extracted by comparing the full-wave results with the tight-binding ones. Figure~\ref{fig:fit} shows the numerically obtained relationship between the dimensionless coupling coefficient $\theta_c L/2$ and the separation $d_c$. As can be seen, the coupling decreases exponentially with increasing separation, consistent with the evanescent coupling nature between the two waveguides. This indicates that it is valid to only consider the couplings between the nearest waveguides in this photonic lattice. Thus, this photonic lattice can be safely mapped to the tight-binding model.

Three coupling values, as highlighted in Fig.~\ref{fig:fit}, are used in the simulation to realize different topological phases. The corresponding waveguide separations are $d_c=22.1$ $\mathrm{\mu m}$ (for $\theta_c L/2 = 0.25 \pi $), $d_c=18.6$ $\mathrm{\mu m}$ (for $\theta_c L/2 = 0.5 \pi$), and $d_c=16.6$ $\mathrm{\mu m}$ (for $\theta_c L/2 =0.75 \pi$). By properly assigning these three coupling values to $J_\alpha$, $J_\beta$, and $M$, we can access all four phases of this model [see Fig.~\ref{fig:evolution}]. 

Next, we construct four finite lattices (each containing eight unit cells) to study the dynamic properties of the four phases. In all simulations, a single-site excitation at the top-left site is adopted, mimicking a realistic experimental scenario. In the trivial phase where no edge modes exist, the light couples into the bulk without edge localization [Fig.~\ref{fig:evolution}(a)]. When $\pm\pi/2$ modes exist, as demonstrated in Fig.~\ref{fig:evolution}(b), a noticeable edge localization can be seen, with a 2$L$-periodic oscillation between the upper and lower leftmost sites. This 2$L$-periodicity can be understood as follows. The initial excitation is a superposition of the $\pm\pi/2$ modes, denoted as $|\psi \rangle = c_1 |\pi /2\rangle + c_2 |-\pi /2\rangle $. After one period, the field becomes
\begin{equation}
U^{\text{SR}} |\psi\rangle = -\mathrm{i} c_1 |\pi /2\rangle + \mathrm{i} c_2 |-\pi /2\rangle,
\end{equation}
which produces a different intensity profile. The original profile can only be restored after two periods:
\begin{equation}
(U^{\text{SR}})^2 |\psi\rangle = -(c_1 |\pi /2\rangle + c_2 |-\pi /2\rangle).
\end{equation}
A similar 2$L$ periodicity happens for the phase with both zero and $\pi$ mode, as shown in Fig.~\ref{fig:evolution}(c). Interestingly, when all types of topological edge modes are present (i.e., zero, $\pi$ and $\pm\pi/2$ modes), the evolution periodicity is extended to 4$L$ [Fig.~\ref{fig:evolution}(d)]. This again can be explained by considering the evolution of an initial state $|\psi \rangle = c_1 |0\rangle + c_2 |\pi \rangle + c_3 |\pi /2\rangle + c_4 |-\pi /2\rangle $:
\begin{align}
    U^{\text{SR}} |\psi \rangle & = \left( c_1 |0\rangle  - c_2 |\pi \rangle  - \mathrm{i} c_3 |\pi /2\rangle + \mathrm{i} c_4 |-\pi /2\rangle \right), \nonumber \\
    {(U^{\text{SR}})}^2 |\psi \rangle & = \left( c_1 |0\rangle + c_2 |\pi \rangle - c_3 |\pi /2\rangle - c_4 |-\pi /2\rangle \right), \nonumber \\ 
    {(U^{\text{SR}})}^3 |\psi \rangle & = \left( c_1 |0\rangle -c_2 |\pi \rangle + \mathrm{i} c_3 |\pi /2\rangle - \mathrm{i} c_4 |-\pi /2\rangle \right), \nonumber \\
    {(U^{\text{SR}})}^4 |\psi \rangle & = \left( c_1 |0\rangle + c_2 |\pi \rangle + c_3 |\pi /2\rangle + c_4 |-\pi /2\rangle \right) \nonumber \\
    & = |\psi \rangle. 
    \label{eq:state_evolution_2}
\end{align}
This 4$L$ periodicity can serve as a hallmark for the existence of $\pi/2$ modes in an experiment.

\textit{Conclusion.}---To conclude, we have proposed a concrete design to realize $\pi/2$ modes in photonics. Our simulations demonstrate that the $\pi/2$ modes can be probed via a simple single-site excitation and intensity measurement at discrete propagation lengths. In practice, this requires fabricating multiple samples with different numbers of driving periods but the same driving protocol, which is expected to be the major experimental challenge. Compared with the previously studied acoustic model~\cite{cheng2022observation}, our photonic system has a much smaller loss and a wider parameter range of coupling strength. Besides, our model can support different types of topological modes and their combinations by just tuning the waveguide separations. This could facilitate further operations on these topological modes, such as braiding.
In future works, it would be interesting to study the system in the quantum regime~\cite{blanco2018topological, wang2019topological, ehrhardt2024topological}, where the unique quasienergies, together with topological protection, can be used to design novel quantum optical devices.

\begin{acknowledgments}
\textit{Acknowledgments.}---This work was supported by the National Natural Science Foundation of China under grant No.~62401491 and the Chinese University of Hong Kong under grants No.~4937205, 4937206 and 4053729. W.Z. acknowledges support from the Start up Funding from the Ocean University of China and the National Natural Science Foundation of China (Grants No.~12404499).
\end{acknowledgments}

\textit{Data availability.}---The data that support the findings of this article are not publicly available upon publication because it is not technically feasible and/or the cost of preparing, depositing, and hosting the data would be prohibitive within the terms of this research project. The data are available from the authors upon reasonable request.

\appendix
\section{Appendix: Numerical details}

We use the split-step Fourier method to solve the Schr\"odinger-like equation numerically (i.e., Eq.~\eqref{eq:schrodinger})~\cite{okamoto2006fundamentals}. We use operator $\mathbf{A}$ to represent the free-space propagation effect and operator $\mathbf{B}$ to represent the waveguiding effect,
\begin{align}
    \mathbf{A} & =  -\frac{1}{2k_0} \nabla_{\perp}^2 , \\ 
    \mathbf{B} & =  - \frac{k_0 \delta n (x,y,z)}{n_0} ,
\end{align}
Then, Eq.~\eqref{eq:schrodinger} reduces to
\begin{equation}
    \mathrm{i} \partial_z \psi =  ( \mathbf{A} + \mathbf{B} ) \psi. 
    \label{eq:schrodinger-1}
\end{equation}
First, we divide the path which light takes into many tiny steps $h$ and assume that operator $\mathbf{B}$ has no $z$-dependence in a step $z_0 < z \leqslant z_0 + h$:
\begin{align} 
    \mathbf{B} =  - \frac{k_0 \delta n (x,y,z_0)}{n_0}  \qquad z_0 \leqslant z < z_0 + h .
\end{align}
Eq.~\eqref{eq:schrodinger-1} can be formally integrated as
\begin{align} 
    \psi(x,y,z_0+h) = \exp[-\mathrm{i} h(\mathbf{A}+\mathbf{B})] \psi(x,y,z_0) .
    \label{eq:wave-evo-0}
\end{align}
Using Baker-Campbell-Hausdorff formula, we approximate Eq.~\eqref{eq:wave-evo-0} as
\begin{align} 
    \psi(x,y,z_0+h) = \exp(-\mathrm{i} \frac{h}{2} \mathbf{B}) \exp(-\mathrm{i} h \mathbf{A}) \nonumber \\ \exp(-\mathrm{i} \frac{h}{2} \mathbf{B}) \psi(x,y,z_0) .
    \label{eq:wave-evo-1}
\end{align}
Then, we need the Fourier transform of $x$ and $y$,
\begin{align} 
    &\Psi(k_x,k_y,z) = F[\psi(x,y,z)] \nonumber \\ &= \int_{-\infty}^{\infty}  \int_{-\infty}^{\infty} \psi(x,y,z) e^{-\mathrm{i} (k_x x + k_y y)} d x d y, \\
    &\psi(x,y,z) = F^{-1}[\Psi(k_x,k_y,z)] \nonumber \\ &= \frac{1}{4\pi^2} \int_{-\infty}^{\infty}  \int_{-\infty}^{\infty} \Psi(k_x,k_y,z) e^{ \mathrm{i} (k_x x + k_y y)} d k_x d k_y, 
\end{align}
where $F$ and $F^{-1}$ are the Fourier transform and the inverse Fourier transform, respectively. Using them to calculate Eq.~\eqref{eq:wave-evo-1}, we get 
\begin{align} 
    & \psi(x,y,z_0+h) = \exp(-\mathrm{i} \frac{h}{2} \mathbf{B})  F^{-1} \nonumber \\  & \left\{ \exp[ -\mathrm{i} \frac{h}{2k_0} (k_x^2 + k_y^2)] F \left[ \exp(-\mathrm{i} \frac{h}{2} \mathbf{B}) \psi(x,y,z_0) \right] \right\} ,
    \label{eq:wave-evo-2}
\end{align}
which describes the evolution of light in the step $[z_0, z_0 + h]$. We then repeat this process to get the whole evolution.


\begin{thebibliography}{27}%
\makeatletter
\providecommand \@ifxundefined [1]{%
 \@ifx{#1\undefined}
}%
\providecommand \@ifnum [1]{%
 \ifnum #1\expandafter \@firstoftwo
 \else \expandafter \@secondoftwo
 \fi
}%
\providecommand \@ifx [1]{%
 \ifx #1\expandafter \@firstoftwo
 \else \expandafter \@secondoftwo
 \fi
}%
\providecommand \natexlab [1]{#1}%
\providecommand \enquote  [1]{``#1''}%
\providecommand \bibnamefont  [1]{#1}%
\providecommand \bibfnamefont [1]{#1}%
\providecommand \citenamefont [1]{#1}%
\providecommand \href@noop [0]{\@secondoftwo}%
\providecommand \href [0]{\begingroup \@sanitize@url \@href}%
\providecommand \@href[1]{\@@startlink{#1}\@@href}%
\providecommand \@@href[1]{\endgroup#1\@@endlink}%
\providecommand \@sanitize@url [0]{\catcode `\\12\catcode `\$12\catcode
  `\&12\catcode `\#12\catcode `\^12\catcode `\_12\catcode `\%12\relax}%
\providecommand \@@startlink[1]{}%
\providecommand \@@endlink[0]{}%
\providecommand \url  [0]{\begingroup\@sanitize@url \@url }%
\providecommand \@url [1]{\endgroup\@href {#1}{\urlprefix }}%
\providecommand \urlprefix  [0]{URL }%
\providecommand \Eprint [0]{\href }%
\providecommand \doibase [0]{https://doi.org/}%
\providecommand \selectlanguage [0]{\@gobble}%
\providecommand \bibinfo  [0]{\@secondoftwo}%
\providecommand \bibfield  [0]{\@secondoftwo}%
\providecommand \translation [1]{[#1]}%
\providecommand \BibitemOpen [0]{}%
\providecommand \bibitemStop [0]{}%
\providecommand \bibitemNoStop [0]{.\EOS\space}%
\providecommand \EOS [0]{\spacefactor3000\relax}%
\providecommand \BibitemShut  [1]{\csname bibitem#1\endcsname}%
\let\auto@bib@innerbib\@empty
\bibitem [{\citenamefont {Oka}\ and\ \citenamefont
  {Kitamura}(2019)}]{oka2019floquet}%
  \BibitemOpen
  \bibfield  {author} {\bibinfo {author} {\bibfnamefont {T.}~\bibnamefont
  {Oka}}\ and\ \bibinfo {author} {\bibfnamefont {S.}~\bibnamefont {Kitamura}},\
  }\bibfield  {title} {\bibinfo {title} {Floquet engineering of quantum
  materials},\ }\href
  {https://doi.org/10.1146/annurev-conmatphys-031218-013423} {\bibfield
  {journal} {\bibinfo  {journal} {Annu. Rev. Condens. Matter Phys.}\ }\textbf
  {\bibinfo {volume} {10}},\ \bibinfo {pages} {387} (\bibinfo {year}
  {2019})}\BibitemShut {NoStop}%
\bibitem [{\citenamefont {Rudner}\ and\ \citenamefont
  {Lindner}(2020)}]{rudner2020band}%
  \BibitemOpen
  \bibfield  {author} {\bibinfo {author} {\bibfnamefont {M.~S.}\ \bibnamefont
  {Rudner}}\ and\ \bibinfo {author} {\bibfnamefont {N.~H.}\ \bibnamefont
  {Lindner}},\ }\bibfield  {title} {\bibinfo {title} {Band structure
  engineering and non-equilibrium dynamics in {F}loquet topological
  insulators},\ }\href {https://doi.org/10.1038/s42254-020-0170-z} {\bibfield
  {journal} {\bibinfo  {journal} {Nat. Rev. Phys.}\ }\textbf {\bibinfo {volume}
  {2}},\ \bibinfo {pages} {229} (\bibinfo {year} {2020})}\BibitemShut {NoStop}%
\bibitem [{\citenamefont {Rudner}\ \emph {et~al.}(2013)\citenamefont {Rudner},
  \citenamefont {Lindner}, \citenamefont {Berg},\ and\ \citenamefont
  {Levin}}]{rudner2013anomalous}%
  \BibitemOpen
  \bibfield  {author} {\bibinfo {author} {\bibfnamefont {M.~S.}\ \bibnamefont
  {Rudner}}, \bibinfo {author} {\bibfnamefont {N.~H.}\ \bibnamefont {Lindner}},
  \bibinfo {author} {\bibfnamefont {E.}~\bibnamefont {Berg}},\ and\ \bibinfo
  {author} {\bibfnamefont {M.}~\bibnamefont {Levin}},\ }\bibfield  {title}
  {\bibinfo {title} {Anomalous edge states and the bulk-edge correspondence for
  periodically driven two-dimensional systems},\ }\href
  {https://doi.org/10.1103/PhysRevX.3.031005} {\bibfield  {journal} {\bibinfo
  {journal} {Phys. Rev. X}\ }\textbf {\bibinfo {volume} {3}},\ \bibinfo {pages}
  {031005} (\bibinfo {year} {2013})}\BibitemShut {NoStop}%
\bibitem [{\citenamefont {Zhang}\ \emph {et~al.}(2021)\citenamefont {Zhang},
  \citenamefont {Delplace},\ and\ \citenamefont {Fleury}}]{zhang2021superior}%
  \BibitemOpen
  \bibfield  {author} {\bibinfo {author} {\bibfnamefont {Z.}~\bibnamefont
  {Zhang}}, \bibinfo {author} {\bibfnamefont {P.}~\bibnamefont {Delplace}},\
  and\ \bibinfo {author} {\bibfnamefont {R.}~\bibnamefont {Fleury}},\
  }\bibfield  {title} {\bibinfo {title} {Superior robustness of anomalous
  non-reciprocal topological edge states},\ }\href
  {https://doi.org/10.1038/s41586-021-03868-7} {\bibfield  {journal} {\bibinfo
  {journal} {Nature}\ }\textbf {\bibinfo {volume} {598}},\ \bibinfo {pages}
  {293} (\bibinfo {year} {2021})}\BibitemShut {NoStop}%
\bibitem [{\citenamefont {Jiang}\ \emph {et~al.}(2011)\citenamefont {Jiang},
  \citenamefont {Kitagawa}, \citenamefont {Alicea}, \citenamefont {Akhmerov},
  \citenamefont {Pekker}, \citenamefont {Refael}, \citenamefont {Cirac},
  \citenamefont {Demler}, \citenamefont {Lukin},\ and\ \citenamefont
  {Zoller}}]{jiang2011majorana}%
  \BibitemOpen
  \bibfield  {author} {\bibinfo {author} {\bibfnamefont {L.}~\bibnamefont
  {Jiang}}, \bibinfo {author} {\bibfnamefont {T.}~\bibnamefont {Kitagawa}},
  \bibinfo {author} {\bibfnamefont {J.}~\bibnamefont {Alicea}}, \bibinfo
  {author} {\bibfnamefont {A.}~\bibnamefont {Akhmerov}}, \bibinfo {author}
  {\bibfnamefont {D.}~\bibnamefont {Pekker}}, \bibinfo {author} {\bibfnamefont
  {G.}~\bibnamefont {Refael}}, \bibinfo {author} {\bibfnamefont {J.~I.}\
  \bibnamefont {Cirac}}, \bibinfo {author} {\bibfnamefont {E.}~\bibnamefont
  {Demler}}, \bibinfo {author} {\bibfnamefont {M.~D.}\ \bibnamefont {Lukin}},\
  and\ \bibinfo {author} {\bibfnamefont {P.}~\bibnamefont {Zoller}},\
  }\bibfield  {title} {\bibinfo {title} {Majorana fermions in equilibrium and
  in driven cold-atom quantum wires},\ }\href
  {https://doi.org/10.1103/PhysRevLett.106.220402} {\bibfield  {journal}
  {\bibinfo  {journal} {Phys. Rev. Lett.}\ }\textbf {\bibinfo {volume} {106}},\
  \bibinfo {pages} {220402} (\bibinfo {year} {2011})}\BibitemShut {NoStop}%
\bibitem [{\citenamefont {Cheng}\ \emph {et~al.}(2019)\citenamefont {Cheng},
  \citenamefont {Pan}, \citenamefont {Wang}, \citenamefont {Zhang},
  \citenamefont {Yu}, \citenamefont {Gover}, \citenamefont {Zhang},
  \citenamefont {Li}, \citenamefont {Zhou},\ and\ \citenamefont
  {Zhu}}]{cheng2019observation}%
  \BibitemOpen
  \bibfield  {author} {\bibinfo {author} {\bibfnamefont {Q.}~\bibnamefont
  {Cheng}}, \bibinfo {author} {\bibfnamefont {Y.}~\bibnamefont {Pan}}, \bibinfo
  {author} {\bibfnamefont {H.}~\bibnamefont {Wang}}, \bibinfo {author}
  {\bibfnamefont {C.}~\bibnamefont {Zhang}}, \bibinfo {author} {\bibfnamefont
  {D.}~\bibnamefont {Yu}}, \bibinfo {author} {\bibfnamefont {A.}~\bibnamefont
  {Gover}}, \bibinfo {author} {\bibfnamefont {H.}~\bibnamefont {Zhang}},
  \bibinfo {author} {\bibfnamefont {T.}~\bibnamefont {Li}}, \bibinfo {author}
  {\bibfnamefont {L.}~\bibnamefont {Zhou}},\ and\ \bibinfo {author}
  {\bibfnamefont {S.}~\bibnamefont {Zhu}},\ }\bibfield  {title} {\bibinfo
  {title} {Observation of anomalous $\pi$ modes in photonic {F}loquet
  engineering},\ }\href {https://doi.org/10.1103/PhysRevLett.122.173901}
  {\bibfield  {journal} {\bibinfo  {journal} {Phys. Rev. Lett.}\ }\textbf
  {\bibinfo {volume} {122}},\ \bibinfo {pages} {173901} (\bibinfo {year}
  {2019})}\BibitemShut {NoStop}%
\bibitem [{\citenamefont {Zhu}\ \emph {et~al.}(2022)\citenamefont {Zhu},
  \citenamefont {Xue}, \citenamefont {Gong}, \citenamefont {Chong},\ and\
  \citenamefont {Zhang}}]{zhu2022time}%
  \BibitemOpen
  \bibfield  {author} {\bibinfo {author} {\bibfnamefont {W.}~\bibnamefont
  {Zhu}}, \bibinfo {author} {\bibfnamefont {H.}~\bibnamefont {Xue}}, \bibinfo
  {author} {\bibfnamefont {J.}~\bibnamefont {Gong}}, \bibinfo {author}
  {\bibfnamefont {Y.}~\bibnamefont {Chong}},\ and\ \bibinfo {author}
  {\bibfnamefont {B.}~\bibnamefont {Zhang}},\ }\bibfield  {title} {\bibinfo
  {title} {Time-periodic corner states from {F}loquet higher-order topology},\
  }\href {https://doi.org/10.1038/s41467-021-27552-6} {\bibfield  {journal}
  {\bibinfo  {journal} {Nat. Commun.}\ }\textbf {\bibinfo {volume} {13}},\
  \bibinfo {pages} {11} (\bibinfo {year} {2022})}\BibitemShut {NoStop}%
\bibitem [{\citenamefont {Ozawa}\ \emph {et~al.}(2019)\citenamefont {Ozawa},
  \citenamefont {Price}, \citenamefont {Amo}, \citenamefont {Goldman},
  \citenamefont {Hafezi}, \citenamefont {Lu}, \citenamefont {Rechtsman},
  \citenamefont {Schuster}, \citenamefont {Simon}, \citenamefont {Zilberberg}
  \emph {et~al.}}]{ozawa2019topological}%
  \BibitemOpen
  \bibfield  {author} {\bibinfo {author} {\bibfnamefont {T.}~\bibnamefont
  {Ozawa}}, \bibinfo {author} {\bibfnamefont {H.~M.}\ \bibnamefont {Price}},
  \bibinfo {author} {\bibfnamefont {A.}~\bibnamefont {Amo}}, \bibinfo {author}
  {\bibfnamefont {N.}~\bibnamefont {Goldman}}, \bibinfo {author} {\bibfnamefont
  {M.}~\bibnamefont {Hafezi}}, \bibinfo {author} {\bibfnamefont
  {L.}~\bibnamefont {Lu}}, \bibinfo {author} {\bibfnamefont {M.~C.}\
  \bibnamefont {Rechtsman}}, \bibinfo {author} {\bibfnamefont {D.}~\bibnamefont
  {Schuster}}, \bibinfo {author} {\bibfnamefont {J.}~\bibnamefont {Simon}},
  \bibinfo {author} {\bibfnamefont {O.}~\bibnamefont {Zilberberg}}, \emph
  {et~al.},\ }\bibfield  {title} {\bibinfo {title} {Topological photonics},\
  }\href {https://doi.org/10.1103/RevModPhys.91.015006} {\bibfield  {journal}
  {\bibinfo  {journal} {Rev. Mod. Phys.}\ }\textbf {\bibinfo {volume} {91}},\
  \bibinfo {pages} {015006} (\bibinfo {year} {2019})}\BibitemShut {NoStop}%
\bibitem [{\citenamefont {Xue}\ \emph {et~al.}(2022)\citenamefont {Xue},
  \citenamefont {Yang},\ and\ \citenamefont {Zhang}}]{xue2022topological}%
  \BibitemOpen
  \bibfield  {author} {\bibinfo {author} {\bibfnamefont {H.}~\bibnamefont
  {Xue}}, \bibinfo {author} {\bibfnamefont {Y.}~\bibnamefont {Yang}},\ and\
  \bibinfo {author} {\bibfnamefont {B.}~\bibnamefont {Zhang}},\ }\bibfield
  {title} {\bibinfo {title} {Topological acoustics},\ }\href
  {https://doi.org/10.1038/s41578-022-00465-6} {\bibfield  {journal} {\bibinfo
  {journal} {Nat. Rev. Mater.}\ }\textbf {\bibinfo {volume} {7}},\ \bibinfo
  {pages} {974} (\bibinfo {year} {2022})}\BibitemShut {NoStop}%
\bibitem [{\citenamefont {Rechtsman}\ \emph {et~al.}(2013)\citenamefont
  {Rechtsman}, \citenamefont {Zeuner}, \citenamefont {Plotnik}, \citenamefont
  {Lumer}, \citenamefont {Podolsky}, \citenamefont {Dreisow}, \citenamefont
  {Nolte}, \citenamefont {Segev},\ and\ \citenamefont
  {Szameit}}]{rechtsman2013photonic}%
  \BibitemOpen
  \bibfield  {author} {\bibinfo {author} {\bibfnamefont {M.~C.}\ \bibnamefont
  {Rechtsman}}, \bibinfo {author} {\bibfnamefont {J.~M.}\ \bibnamefont
  {Zeuner}}, \bibinfo {author} {\bibfnamefont {Y.}~\bibnamefont {Plotnik}},
  \bibinfo {author} {\bibfnamefont {Y.}~\bibnamefont {Lumer}}, \bibinfo
  {author} {\bibfnamefont {D.}~\bibnamefont {Podolsky}}, \bibinfo {author}
  {\bibfnamefont {F.}~\bibnamefont {Dreisow}}, \bibinfo {author} {\bibfnamefont
  {S.}~\bibnamefont {Nolte}}, \bibinfo {author} {\bibfnamefont
  {M.}~\bibnamefont {Segev}},\ and\ \bibinfo {author} {\bibfnamefont
  {A.}~\bibnamefont {Szameit}},\ }\bibfield  {title} {\bibinfo {title}
  {Photonic {F}loquet topological insulators},\ }\href
  {https://doi.org/10.1038/nature12066} {\bibfield  {journal} {\bibinfo
  {journal} {Nature}\ }\textbf {\bibinfo {volume} {496}},\ \bibinfo {pages}
  {196} (\bibinfo {year} {2013})}\BibitemShut {NoStop}%
\bibitem [{\citenamefont {Gao}\ \emph {et~al.}(2016)\citenamefont {Gao},
  \citenamefont {Gao}, \citenamefont {Shi}, \citenamefont {Yang}, \citenamefont
  {Lin}, \citenamefont {Xu}, \citenamefont {Joannopoulos}, \citenamefont
  {Solja{\v{c}}i{\'c}}, \citenamefont {Chen}, \citenamefont {Lu} \emph
  {et~al.}}]{gao2016probing}%
  \BibitemOpen
  \bibfield  {author} {\bibinfo {author} {\bibfnamefont {F.}~\bibnamefont
  {Gao}}, \bibinfo {author} {\bibfnamefont {Z.}~\bibnamefont {Gao}}, \bibinfo
  {author} {\bibfnamefont {X.}~\bibnamefont {Shi}}, \bibinfo {author}
  {\bibfnamefont {Z.}~\bibnamefont {Yang}}, \bibinfo {author} {\bibfnamefont
  {X.}~\bibnamefont {Lin}}, \bibinfo {author} {\bibfnamefont {H.}~\bibnamefont
  {Xu}}, \bibinfo {author} {\bibfnamefont {J.~D.}\ \bibnamefont
  {Joannopoulos}}, \bibinfo {author} {\bibfnamefont {M.}~\bibnamefont
  {Solja{\v{c}}i{\'c}}}, \bibinfo {author} {\bibfnamefont {H.}~\bibnamefont
  {Chen}}, \bibinfo {author} {\bibfnamefont {L.}~\bibnamefont {Lu}}, \emph
  {et~al.},\ }\bibfield  {title} {\bibinfo {title} {Probing topological
  protection using a designer surface plasmon structure},\ }\href
  {https://doi.org/10.1038/ncomms11619} {\bibfield  {journal} {\bibinfo
  {journal} {Nat. Commun.}\ }\textbf {\bibinfo {volume} {7}},\ \bibinfo {pages}
  {11619} (\bibinfo {year} {2016})}\BibitemShut {NoStop}%
\bibitem [{\citenamefont {Peng}\ \emph {et~al.}(2016)\citenamefont {Peng},
  \citenamefont {Qin}, \citenamefont {Zhao}, \citenamefont {Shen},
  \citenamefont {Xu}, \citenamefont {Bao}, \citenamefont {Jia},\ and\
  \citenamefont {Zhu}}]{peng2016experimental}%
  \BibitemOpen
  \bibfield  {author} {\bibinfo {author} {\bibfnamefont {Y.-G.}\ \bibnamefont
  {Peng}}, \bibinfo {author} {\bibfnamefont {C.-Z.}\ \bibnamefont {Qin}},
  \bibinfo {author} {\bibfnamefont {D.-G.}\ \bibnamefont {Zhao}}, \bibinfo
  {author} {\bibfnamefont {Y.-X.}\ \bibnamefont {Shen}}, \bibinfo {author}
  {\bibfnamefont {X.-Y.}\ \bibnamefont {Xu}}, \bibinfo {author} {\bibfnamefont
  {M.}~\bibnamefont {Bao}}, \bibinfo {author} {\bibfnamefont {H.}~\bibnamefont
  {Jia}},\ and\ \bibinfo {author} {\bibfnamefont {X.-F.}\ \bibnamefont {Zhu}},\
  }\bibfield  {title} {\bibinfo {title} {Experimental demonstration of
  anomalous {F}loquet topological insulator for sound},\ }\href
  {https://doi.org/10.1038/ncomms13368} {\bibfield  {journal} {\bibinfo
  {journal} {Nat. Commun.}\ }\textbf {\bibinfo {volume} {7}},\ \bibinfo {pages}
  {13368} (\bibinfo {year} {2016})}\BibitemShut {NoStop}%
\bibitem [{\citenamefont {Maczewsky}\ \emph {et~al.}(2017)\citenamefont
  {Maczewsky}, \citenamefont {Zeuner}, \citenamefont {Nolte},\ and\
  \citenamefont {Szameit}}]{maczewsky2017observation}%
  \BibitemOpen
  \bibfield  {author} {\bibinfo {author} {\bibfnamefont {L.~J.}\ \bibnamefont
  {Maczewsky}}, \bibinfo {author} {\bibfnamefont {J.~M.}\ \bibnamefont
  {Zeuner}}, \bibinfo {author} {\bibfnamefont {S.}~\bibnamefont {Nolte}},\ and\
  \bibinfo {author} {\bibfnamefont {A.}~\bibnamefont {Szameit}},\ }\bibfield
  {title} {\bibinfo {title} {Observation of photonic anomalous {F}loquet
  topological insulators},\ }\href {https://doi.org/10.1038/ncomms13756}
  {\bibfield  {journal} {\bibinfo  {journal} {Nat. Commun.}\ }\textbf {\bibinfo
  {volume} {8}},\ \bibinfo {pages} {13756} (\bibinfo {year}
  {2017})}\BibitemShut {NoStop}%
\bibitem [{\citenamefont {Mukherjee}\ \emph {et~al.}(2017)\citenamefont
  {Mukherjee}, \citenamefont {Spracklen}, \citenamefont {Valiente},
  \citenamefont {Andersson}, \citenamefont {{\"O}hberg}, \citenamefont
  {Goldman},\ and\ \citenamefont {Thomson}}]{mukherjee2017experimental}%
  \BibitemOpen
  \bibfield  {author} {\bibinfo {author} {\bibfnamefont {S.}~\bibnamefont
  {Mukherjee}}, \bibinfo {author} {\bibfnamefont {A.}~\bibnamefont
  {Spracklen}}, \bibinfo {author} {\bibfnamefont {M.}~\bibnamefont {Valiente}},
  \bibinfo {author} {\bibfnamefont {E.}~\bibnamefont {Andersson}}, \bibinfo
  {author} {\bibfnamefont {P.}~\bibnamefont {{\"O}hberg}}, \bibinfo {author}
  {\bibfnamefont {N.}~\bibnamefont {Goldman}},\ and\ \bibinfo {author}
  {\bibfnamefont {R.~R.}\ \bibnamefont {Thomson}},\ }\bibfield  {title}
  {\bibinfo {title} {Experimental observation of anomalous topological edge
  modes in a slowly driven photonic lattice},\ }\href
  {https://doi.org/10.1038/ncomms13918} {\bibfield  {journal} {\bibinfo
  {journal} {Nat. Commun.}\ }\textbf {\bibinfo {volume} {8}},\ \bibinfo {pages}
  {13918} (\bibinfo {year} {2017})}\BibitemShut {NoStop}%
\bibitem [{\citenamefont {Maczewsky}\ \emph {et~al.}(2020)\citenamefont
  {Maczewsky}, \citenamefont {H{\"o}ckendorf}, \citenamefont {Kremer},
  \citenamefont {Biesenthal}, \citenamefont {Heinrich}, \citenamefont
  {Alvermann}, \citenamefont {Fehske},\ and\ \citenamefont
  {Szameit}}]{maczewsky2020fermionic}%
  \BibitemOpen
  \bibfield  {author} {\bibinfo {author} {\bibfnamefont {L.~J.}\ \bibnamefont
  {Maczewsky}}, \bibinfo {author} {\bibfnamefont {B.}~\bibnamefont
  {H{\"o}ckendorf}}, \bibinfo {author} {\bibfnamefont {M.}~\bibnamefont
  {Kremer}}, \bibinfo {author} {\bibfnamefont {T.}~\bibnamefont {Biesenthal}},
  \bibinfo {author} {\bibfnamefont {M.}~\bibnamefont {Heinrich}}, \bibinfo
  {author} {\bibfnamefont {A.}~\bibnamefont {Alvermann}}, \bibinfo {author}
  {\bibfnamefont {H.}~\bibnamefont {Fehske}},\ and\ \bibinfo {author}
  {\bibfnamefont {A.}~\bibnamefont {Szameit}},\ }\bibfield  {title} {\bibinfo
  {title} {Fermionic time-reversal symmetry in a photonic topological
  insulator},\ }\href {https://doi.org/10.1038/s41563-020-0641-8} {\bibfield
  {journal} {\bibinfo  {journal} {Nat. Mater.}\ }\textbf {\bibinfo {volume}
  {19}},\ \bibinfo {pages} {855} (\bibinfo {year} {2020})}\BibitemShut
  {NoStop}%
\bibitem [{\citenamefont {Wu}\ \emph {et~al.}(2021)\citenamefont {Wu},
  \citenamefont {Song}, \citenamefont {Gao}, \citenamefont {Chen},
  \citenamefont {Zhu},\ and\ \citenamefont {Li}}]{wu2021floquet}%
  \BibitemOpen
  \bibfield  {author} {\bibinfo {author} {\bibfnamefont {S.}~\bibnamefont
  {Wu}}, \bibinfo {author} {\bibfnamefont {W.}~\bibnamefont {Song}}, \bibinfo
  {author} {\bibfnamefont {S.}~\bibnamefont {Gao}}, \bibinfo {author}
  {\bibfnamefont {Y.}~\bibnamefont {Chen}}, \bibinfo {author} {\bibfnamefont
  {S.}~\bibnamefont {Zhu}},\ and\ \bibinfo {author} {\bibfnamefont
  {T.}~\bibnamefont {Li}},\ }\bibfield  {title} {\bibinfo {title} {Floquet
  $\pi$ mode engineering in non-{H}ermitian waveguide lattices},\ }\href
  {https://doi.org/10.1103/PhysRevResearch.3.023211} {\bibfield  {journal}
  {\bibinfo  {journal} {Phys. Rev. Research}\ }\textbf {\bibinfo {volume}
  {3}},\ \bibinfo {pages} {023211} (\bibinfo {year} {2021})}\BibitemShut
  {NoStop}%
\bibitem [{\citenamefont {Sidorenko}\ \emph {et~al.}(2022)\citenamefont
  {Sidorenko}, \citenamefont {Fedorova}, \citenamefont {Abouelela},
  \citenamefont {Kroha},\ and\ \citenamefont {Linden}}]{sidorenko2022real}%
  \BibitemOpen
  \bibfield  {author} {\bibinfo {author} {\bibfnamefont {A.}~\bibnamefont
  {Sidorenko}}, \bibinfo {author} {\bibfnamefont {Z.}~\bibnamefont {Fedorova}},
  \bibinfo {author} {\bibfnamefont {A.}~\bibnamefont {Abouelela}}, \bibinfo
  {author} {\bibfnamefont {J.}~\bibnamefont {Kroha}},\ and\ \bibinfo {author}
  {\bibfnamefont {S.}~\bibnamefont {Linden}},\ }\bibfield  {title} {\bibinfo
  {title} {Real-and {F}ourier-space observation of the anomalous $\pi$ mode in
  {F}loquet engineered plasmonic waveguide arrays},\ }\href
  {https://doi.org/10.1103/PhysRevResearch.4.033184} {\bibfield  {journal}
  {\bibinfo  {journal} {Phys. Rev. Research}\ }\textbf {\bibinfo {volume}
  {4}},\ \bibinfo {pages} {033184} (\bibinfo {year} {2022})}\BibitemShut
  {NoStop}%
\bibitem [{\citenamefont {Cheng}\ \emph {et~al.}(2022)\citenamefont {Cheng},
  \citenamefont {Bomantara}, \citenamefont {Xue}, \citenamefont {Zhu},
  \citenamefont {Gong},\ and\ \citenamefont {Zhang}}]{cheng2022observation}%
  \BibitemOpen
  \bibfield  {author} {\bibinfo {author} {\bibfnamefont {Z.}~\bibnamefont
  {Cheng}}, \bibinfo {author} {\bibfnamefont {R.~W.}\ \bibnamefont
  {Bomantara}}, \bibinfo {author} {\bibfnamefont {H.}~\bibnamefont {Xue}},
  \bibinfo {author} {\bibfnamefont {W.}~\bibnamefont {Zhu}}, \bibinfo {author}
  {\bibfnamefont {J.}~\bibnamefont {Gong}},\ and\ \bibinfo {author}
  {\bibfnamefont {B.}~\bibnamefont {Zhang}},\ }\bibfield  {title} {\bibinfo
  {title} {Observation of $\pi$/2 modes in an acoustic {F}loquet system},\
  }\href {https://doi.org/10.1103/PhysRevLett.129.254301} {\bibfield  {journal}
  {\bibinfo  {journal} {Phys. Rev. Lett.}\ }\textbf {\bibinfo {volume} {129}},\
  \bibinfo {pages} {254301} (\bibinfo {year} {2022})}\BibitemShut {NoStop}%
\bibitem [{\citenamefont {Sreejith}\ \emph {et~al.}(2016)\citenamefont
  {Sreejith}, \citenamefont {Lazarides},\ and\ \citenamefont
  {Moessner}}]{sreejith2016parafermion}%
  \BibitemOpen
  \bibfield  {author} {\bibinfo {author} {\bibfnamefont {G.}~\bibnamefont
  {Sreejith}}, \bibinfo {author} {\bibfnamefont {A.}~\bibnamefont
  {Lazarides}},\ and\ \bibinfo {author} {\bibfnamefont {R.}~\bibnamefont
  {Moessner}},\ }\bibfield  {title} {\bibinfo {title} {Parafermion chain with 2
  $\pi$/k {F}loquet edge modes},\ }\href
  {https://doi.org/10.1103/PhysRevB.94.045127} {\bibfield  {journal} {\bibinfo
  {journal} {Phys. Rev. B}\ }\textbf {\bibinfo {volume} {94}},\ \bibinfo
  {pages} {045127} (\bibinfo {year} {2016})}\BibitemShut {NoStop}%
\bibitem [{\citenamefont {Bomantara}(2021)}]{bomantara2021z}%
  \BibitemOpen
  \bibfield  {author} {\bibinfo {author} {\bibfnamefont {R.~W.}\ \bibnamefont
  {Bomantara}},\ }\bibfield  {title} {\bibinfo {title} {${Z}_{4}$
  parafermion$\pm$$\pi$/2 modes in an interacting periodically driven
  superconducting chain},\ }\href
  {https://doi.org/10.1103/PhysRevB.104.L121410} {\bibfield  {journal}
  {\bibinfo  {journal} {Phys. Rev. B}\ }\textbf {\bibinfo {volume} {104}},\
  \bibinfo {pages} {L121410} (\bibinfo {year} {2021})}\BibitemShut {NoStop}%
\bibitem [{\citenamefont {Bomantara}(2022)}]{bomantara2022square}%
  \BibitemOpen
  \bibfield  {author} {\bibinfo {author} {\bibfnamefont {R.~W.}\ \bibnamefont
  {Bomantara}},\ }\bibfield  {title} {\bibinfo {title} {Square-root {F}loquet
  topological phases and time crystals},\ }\href
  {https://doi.org/10.1103/PhysRevB.106.L060305} {\bibfield  {journal}
  {\bibinfo  {journal} {Phys. Rev. B}\ }\textbf {\bibinfo {volume} {106}},\
  \bibinfo {pages} {L060305} (\bibinfo {year} {2022})}\BibitemShut {NoStop}%
\bibitem [{\citenamefont {Kang}\ \emph {et~al.}(2023)\citenamefont {Kang},
  \citenamefont {Wei}, \citenamefont {Zhang},\ and\ \citenamefont
  {Dong}}]{kang2023topological}%
  \BibitemOpen
  \bibfield  {author} {\bibinfo {author} {\bibfnamefont {J.}~\bibnamefont
  {Kang}}, \bibinfo {author} {\bibfnamefont {R.}~\bibnamefont {Wei}}, \bibinfo
  {author} {\bibfnamefont {Q.}~\bibnamefont {Zhang}},\ and\ \bibinfo {author}
  {\bibfnamefont {G.}~\bibnamefont {Dong}},\ }\bibfield  {title} {\bibinfo
  {title} {Topological photonic states in waveguide arrays},\ }\href
  {https://doi.org/10.1002/apxr.202200053} {\bibfield  {journal} {\bibinfo
  {journal} {Adv. Phys. Research}\ }\textbf {\bibinfo {volume} {2}},\ \bibinfo
  {pages} {2200053} (\bibinfo {year} {2023})}\BibitemShut {NoStop}%
\bibitem [{\citenamefont {Arkinstall}\ \emph {et~al.}(2017)\citenamefont
  {Arkinstall}, \citenamefont {Teimourpour}, \citenamefont {Feng},
  \citenamefont {El-Ganainy},\ and\ \citenamefont
  {Schomerus}}]{arkinstall2017topological}%
  \BibitemOpen
  \bibfield  {author} {\bibinfo {author} {\bibfnamefont {J.}~\bibnamefont
  {Arkinstall}}, \bibinfo {author} {\bibfnamefont {M.}~\bibnamefont
  {Teimourpour}}, \bibinfo {author} {\bibfnamefont {L.}~\bibnamefont {Feng}},
  \bibinfo {author} {\bibfnamefont {R.}~\bibnamefont {El-Ganainy}},\ and\
  \bibinfo {author} {\bibfnamefont {H.}~\bibnamefont {Schomerus}},\ }\bibfield
  {title} {\bibinfo {title} {Topological tight-binding models from nontrivial
  square roots},\ }\href {https://doi.org/10.1103/PhysRevB.95.165109}
  {\bibfield  {journal} {\bibinfo  {journal} {Phys. Rev. B}\ }\textbf {\bibinfo
  {volume} {95}},\ \bibinfo {pages} {165109} (\bibinfo {year}
  {2017})}\BibitemShut {NoStop}%
\bibitem [{\citenamefont {Okamoto}(2006)}]{okamoto2006fundamentals}%
  \BibitemOpen
  \bibfield  {author} {\bibinfo {author} {\bibfnamefont {K.}~\bibnamefont
  {Okamoto}},\ }\href@noop {} {\emph {\bibinfo {title} {Fundamentals of Optical
  Waveguides}}},\ \bibinfo {edition} {2nd}\ ed.\ (\bibinfo  {publisher}
  {Elsevier},\ \bibinfo {address} {Amsterdam Boston},\ \bibinfo {year}
  {2006})\BibitemShut {NoStop}%
\bibitem [{\citenamefont {Blanco-Redondo}\ \emph {et~al.}(2018)\citenamefont
  {Blanco-Redondo}, \citenamefont {Bell}, \citenamefont {Oren}, \citenamefont
  {Eggleton},\ and\ \citenamefont {Segev}}]{blanco2018topological}%
  \BibitemOpen
  \bibfield  {author} {\bibinfo {author} {\bibfnamefont {A.}~\bibnamefont
  {Blanco-Redondo}}, \bibinfo {author} {\bibfnamefont {B.}~\bibnamefont
  {Bell}}, \bibinfo {author} {\bibfnamefont {D.}~\bibnamefont {Oren}}, \bibinfo
  {author} {\bibfnamefont {B.~J.}\ \bibnamefont {Eggleton}},\ and\ \bibinfo
  {author} {\bibfnamefont {M.}~\bibnamefont {Segev}},\ }\bibfield  {title}
  {\bibinfo {title} {Topological protection of biphoton states},\ }\href
  {https://doi.org/10.1126/science.aau4296} {\bibfield  {journal} {\bibinfo
  {journal} {Science}\ }\textbf {\bibinfo {volume} {362}},\ \bibinfo {pages}
  {568} (\bibinfo {year} {2018})}\BibitemShut {NoStop}%
\bibitem [{\citenamefont {Wang}\ \emph {et~al.}(2019)\citenamefont {Wang},
  \citenamefont {Pang}, \citenamefont {Lu}, \citenamefont {Gao}, \citenamefont
  {Chang}, \citenamefont {Qiao}, \citenamefont {Jiao}, \citenamefont {Tang},\
  and\ \citenamefont {Jin}}]{wang2019topological}%
  \BibitemOpen
  \bibfield  {author} {\bibinfo {author} {\bibfnamefont {Y.}~\bibnamefont
  {Wang}}, \bibinfo {author} {\bibfnamefont {X.-L.}\ \bibnamefont {Pang}},
  \bibinfo {author} {\bibfnamefont {Y.-H.}\ \bibnamefont {Lu}}, \bibinfo
  {author} {\bibfnamefont {J.}~\bibnamefont {Gao}}, \bibinfo {author}
  {\bibfnamefont {Y.-J.}\ \bibnamefont {Chang}}, \bibinfo {author}
  {\bibfnamefont {L.-F.}\ \bibnamefont {Qiao}}, \bibinfo {author}
  {\bibfnamefont {Z.-Q.}\ \bibnamefont {Jiao}}, \bibinfo {author}
  {\bibfnamefont {H.}~\bibnamefont {Tang}},\ and\ \bibinfo {author}
  {\bibfnamefont {X.-M.}\ \bibnamefont {Jin}},\ }\bibfield  {title} {\bibinfo
  {title} {Topological protection of two-photon quantum correlation on a
  photonic chip},\ }\href {https://doi.org/10.1364/OPTICA.6.000955} {\bibfield
  {journal} {\bibinfo  {journal} {Optica}\ }\textbf {\bibinfo {volume} {6}},\
  \bibinfo {pages} {955} (\bibinfo {year} {2019})}\BibitemShut {NoStop}%
\bibitem [{\citenamefont {Ehrhardt}\ \emph {et~al.}(2024)\citenamefont
  {Ehrhardt}, \citenamefont {Dittel}, \citenamefont {Heinrich},\ and\
  \citenamefont {Szameit}}]{ehrhardt2024topological}%
  \BibitemOpen
  \bibfield  {author} {\bibinfo {author} {\bibfnamefont {M.}~\bibnamefont
  {Ehrhardt}}, \bibinfo {author} {\bibfnamefont {C.}~\bibnamefont {Dittel}},
  \bibinfo {author} {\bibfnamefont {M.}~\bibnamefont {Heinrich}},\ and\
  \bibinfo {author} {\bibfnamefont {A.}~\bibnamefont {Szameit}},\ }\bibfield
  {title} {\bibinfo {title} {Topological {H}ong-{O}u-{M}andel interference},\
  }\href {https://doi.org/10.1126/science.ado8192} {\bibfield  {journal}
  {\bibinfo  {journal} {Science}\ }\textbf {\bibinfo {volume} {384}},\ \bibinfo
  {pages} {1340} (\bibinfo {year} {2024})}\BibitemShut {NoStop}%
\end{thebibliography}
\end{document}